\documentclass[11pt,preprint]{aastex}

\begin{document}
\shorttitle{V4046 Sgr Disk Kinematics}
\shortauthors{Rosenfeld et al.}

\title{A Disk-based Dynamical Mass Estimate for the Young Binary V4046 Sgr}

\author{Katherine A. Rosenfeld\altaffilmark{1}, 
Sean M. Andrews\altaffilmark{1}, 
David J. Wilner\altaffilmark{1}, 
\& H. C. Stempels\altaffilmark{2}}
\altaffiltext{1}{Harvard-Smithsonian Center for Astrophysics, 60 Garden Street, Cambridge, MA 02138}
\altaffiltext{2}{Department Physics and Astronomy, Uppsala University, Box 516, 751 20 Uppsala, Sweden}

\begin{abstract}
We present sensitive, arcsecond-resolution Submillimeter Array observations of
the $^{12}$CO $J$=2$-$1 line emission from the circumstellar disk orbiting the 
double-lined spectroscopic binary star V4046 Sgr.  Based on a simple model of 
the disk structure, we use a novel Monte Carlo Markov Chain technique to 
extract the Keplerian velocity field of the disk from these data and estimate 
the total mass of the central binary.  Assuming the distance inferred from 
kinematic parallax measurements in the literature ($d \approx 73$\,pc), we 
determine a total stellar mass $M_{\ast} = 1.75^{+0.09}_{-0.06}$\,M$_{\odot}$ 
and a disk inclination $i_d = 33\fdg5^{+0.7}_{-1.4}$ from face-on.  These 
measurements are in excellent agreement with independent dynamical constraints 
made from multi-epoch monitoring of the stellar radial velocities, confirming 
the absolute accuracy of this precise ($\sim$few percent uncertainties) 
disk-based method for estimating stellar masses and reaffirming previous 
assertions that the disk and binary orbital planes are well aligned (with 
$|i_d - i_{\ast}| \approx 0.1\pm1\degr$).  Using these results as a reference, 
we demonstrate that various pre-main sequence evolution models make consistent 
and accurate predictions for the masses of the individual components of the 
binary, and uniformly imply an advanced age of $\sim$5-30\,Myr.  Taken 
together, these results verify that V4046 Sgr is one of the precious few nearby 
and relatively evolved pre-main sequence systems that still hosts a gas-rich 
accretion disk.
\end{abstract}

\section{Introduction} \label{sec:intro}

Mass is the fundamental property that sets the evolutionary path of a star.  
The masses of young stars are of particular interest in many astrophysical 
problems: they provide unique information about the star formation process 
\citep[e.g., accretion histories, the initial mass function; see][]{bastian10} 
and are thought to have a substantial influence on the evolution of their 
circumstellar material, and therefore the efficiency of the planet formation 
process \citep[e.g., see][]{alibert11}.  Unfortunately, measurements of 
pre-main sequence (pre-MS) star masses are difficult and accordingly rare.  
Unlike their more evolved counterparts, the location of a young star in the 
Hertzsprung-Russell (HR) diagram does not provide a robust estimate of 
$M_{\ast}$ \citep{hillenbrand04}.  The theoretical models for stellar evolution 
at these early stages are plagued with uncertainties related to rotation 
\citep{siess97a,mendes99}, accretion \citep{siess97b,baraffe09,baraffe10}, 
magnetic fields \citep{dantona00}, atmosphere properties \citep[e.g., 
convection, opacities;][]{baraffe02}, and unknown initial conditions.  
Ultimately, a more nuanced understanding of star and planet formation requires 
pre-MS evolution models that are empirically calibrated with direct, 
independent, and accurate $M_{\ast}$ measurements.  

The only {\it direct} methods available for measuring stellar masses are based 
on orbital dynamics.  In sufficiently close pre-MS binary star systems, 
$M_{\ast}$ can be estimated from the stellar orbits using multi-epoch radial 
velocity (RV) measurements \citep[e.g.,][]{mathieu89,mathieu91,mathieu97} 
and/or long-term astrometric monitoring 
\citep{tamazian02,schaefer03,schaefer06,duchene06}.  For a double-lined 
spectroscopic binary (SB2), the RV method provides a robust estimate of 
$M_{\ast} (\sin{i_{\ast}})^3$ for each component, where $i_{\ast}$ is the 
inclination angle of the orbit projected on the sky.  For ``visual" binaries, 
the astrometric monitoring technique offers a constraint on the quantity 
$M_{\rm tot}d^3$, where $M_{\rm tot}$ is the sum of the stellar masses and $d$ 
is the distance to the binary.  Generally, the stellar mass estimates from 
these methods are inherently uncertain due to their strong dependences on the 
unknown values of $i_{\ast}$ or $d$.  The \{$M_{\ast}$, $i_{\ast}$\} degeneracy 
is broken for the special case of an eclipsing SB2, but few pre-MS systems with 
such favorable orientations are known \citep[e.g.,][and references therein]{stassun04,morales12}.  In a subset of ideal cases, the RV and 
astrometric techniques can be combined to alleviate the uncertainties related 
to \{$i_{\ast}$, $d$\} and extract accurate $M_{\ast}$ values 
\citep[e.g.,][]{steffen01,schaefer08,boden05,boden07,boden09,boden12}.

These standard methods are only applicable for binary stars with a narrow range 
of orbital separations.  Alternatively, for any {\it isolated} young star with 
a circumstellar disk, $M_{\ast}$ can be determined from a single 
millimeter-wave interferometric observation of an optically thick emission line 
(with a linear dependence on $d$).  This latter technique relies on modeling 
the spatially and spectrally resolved Keplerian rotation curve of the molecular 
gas disk that orbits the young star 
\citep{koerner93,dutrey94,dutrey98,simon00}.  While this method has 
extraordinary value in its more general applicability, it has only been 
successfully employed for small samples.  Attempts to expand its reach have 
been frustrated by molecular cloud contamination and observational limitations 
in resolution and sensitivity.  Moreover, a reconstruction of the disk velocity 
field necessarily involves fitting such data with a relatively complicated 
model of the disk structure \citep{beckwith93}.  Given that added complexity, 
there is naturally some concern about the absolute accuracy of the $M_{\ast}$ 
estimates from this method \citep[e.g.,][]{gennaro12}, despite the impressive 
formal precision of the measurements \citep[$\sim$2-3\%; e.g.,][]{pietu07}.  

The young binary V4046 Sgr provides a rare opportunity to benchmark the disk 
kinematics method for estimating $M_{\ast}$ against the more traditional RV 
technique for a SB2 system.  V4046 Sgr is a nearly equal mass ($q \approx 
0.94$) pair of solar-type pre-MS stars in a circular ($e \le 0.01$), 
non-eclipsing orbit with a 2.4 day period \citep[$a \sin{i_{\ast}} \approx 
5.1$\,R$_{\odot}$;][]{byrne86,quast00,stempels04}.  The system is completely 
isolated from any known molecular clouds and has been kinematically associated 
with the $\sim$8-20\,Myr-old $\beta$ Pic moving group; a moving cluster 
analysis suggests it is relatively nearby, $d = 73$\,pc \citep{torres06}.  
Despite its advanced age, V4046 Sgr hosts a large and massive circumbinary disk 
that exhibits a rich molecular emission line spectrum 
\citep{kastner08,rodriguez10,oberg11}.  Since the binary orbit is tight and 
circular, it has no dynamical impact on the disk structure outside a radius of 
$\sim$0.1\,AU \citep[e.g., see][]{artymowicz94}.  With its central SB2 host, 
rare proximity to the Sun, lack of molecular cloud contamination, and 
intrinsically bright, spatially extended line emission, the V4046 Sgr disk is 
an ideal target to assess the accuracy of the disk kinematics technique for 
measuring $M_{\ast}$.

In this article, we build on some initial work by \citet{rodriguez10} and use 
high-quality spatially and spectrally resolved observations of the CO $J$=2$-$1 
emission line to measure the velocity field of the V4046 Sgr disk and extract 
the total mass of the close binary at its center.  Our millimeter-wave 
observations with the Submillimeter Array (SMA) and data calibration procedures 
are described in \S \ref{sec:observations}.  A detailed overview of the 
modeling analysis is provided in \S \ref{sec:modeling}.  The modeling results 
are presented and compared with the complementary RV analysis of the central 
SB2 by \citet{stempels12} in \S \ref{sec:results}.  These results are discussed 
in the context of the V4046 Sgr system in particular, pre-MS evolution models 
more generally, and future prospects for $M_{\ast}$ estimates from the disk 
kinematics method in \S \ref{sec:discuss}.  Some key conclusions from this work 
are summarized in \S \ref{sec:summary}.

\section{Observations and Data Reduction} \label{sec:observations}

The V4046 Sgr circumbinary disk was observed at 225\,GHz (1.3\,mm) with the 
Submillimeter Array \citep[SMA;][]{ho04} on four occasions starting in 2009, 
using each of the available antenna configurations: sub-compact (baseline 
lengths of 9-25\,m; 2011 Mar 18), compact (16-70\,m; 2009 Apr 25), extended 
(28-226\,m; 2009 Feb 23), and very extended (68-509\,m; 2011 Sep 4).  The SMA 
double sideband receivers were tuned to simultaneously observe the $J$=2$-$1 
transitions of $^{12}$CO, $^{13}$CO, and C$^{18}$O at 230.538, 220.399, and 
219.560\,GHz, respectively, and the adjacent dust continuum.  The correlator 
was configured to place those emission lines in separate 104\,MHz spectral 
chunks and sample them finely with 512 channels per chunk, corresponding to a 
native velocity resolution of $\sim$0.25\,km s$^{-1}$.  The continuum was 
observed with a more coarse frequency sampling (in 3.25\,MHz channels), with a 
total bandwidth of 1.6 and 3.6\,GHz in 2009 and 2011, respectively.  The 
observations cycled between V4046 Sgr and the quasars J1924-292 (15\degr\ away) 
and J1733-130 (22\degr\ from V4046 Sgr, 30\degr\ from J1924-292) with a total 
loop time of 10-20 minutes.  The bright quasars 3C 84 and 3C 454.3 were also 
observed as bandpass calibrators, along with Uranus, Callisto, and Ceres for
use in determining the absolute scaling of the amplitudes.  All of the data 
were collected in outstanding weather conditions for this observing frequency, 
with an atmospheric zenith optical depth of only $\sim$0.05 (corresponding to a 
precipitable water vapor level of $\sim$1.0\,mm).  

\begin{figure}[t!]
\epsscale{1.0}
\plotone{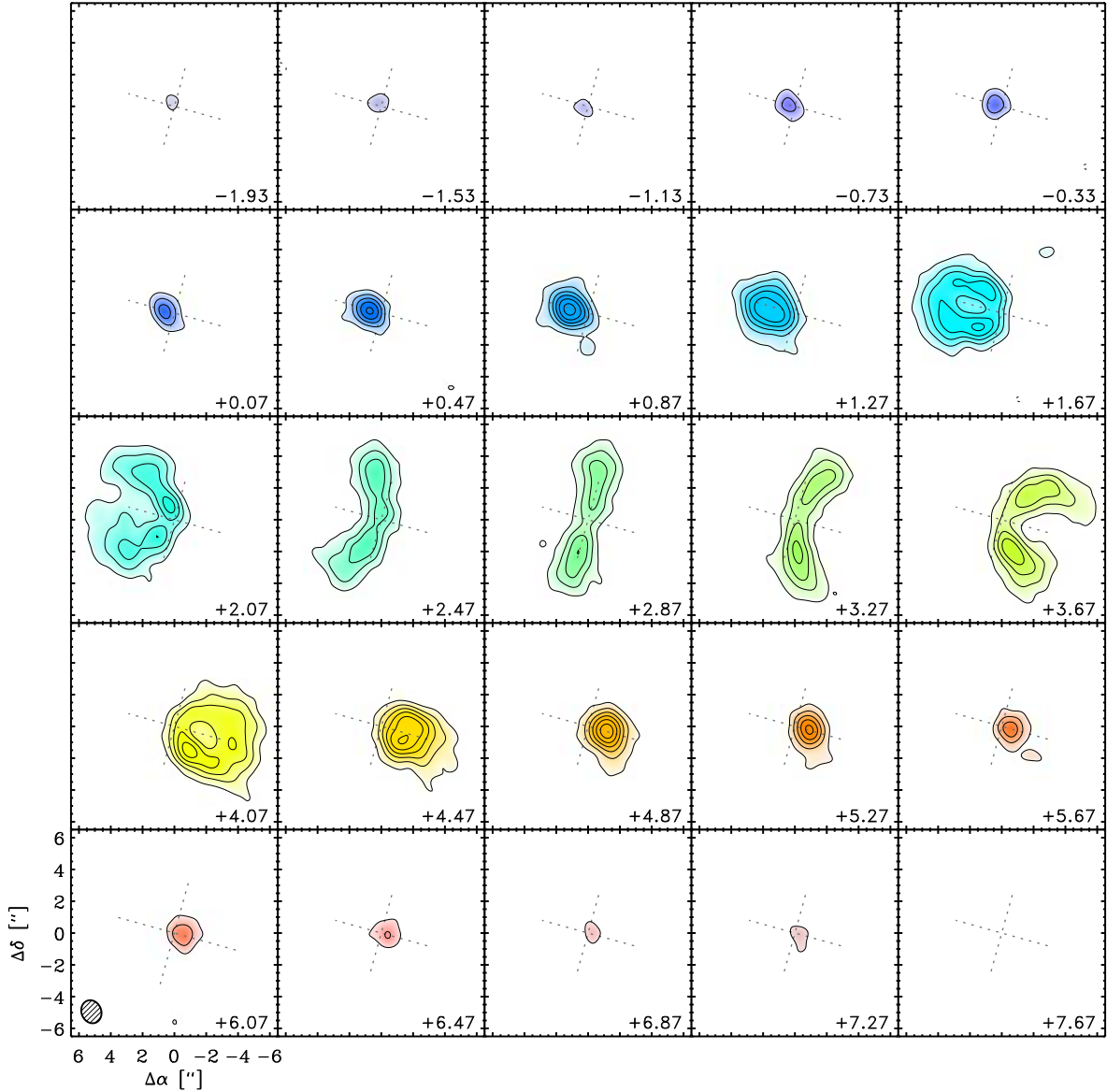}
\caption{Naturally weighted channel maps of $^{12}$CO $J$=2$-$1 emission from 
the V4046 Sgr disk.  The synthesized beam dimensions are marked in the bottom 
left.  In each 0.4\,km s$^{-1}$-wide channel (LSR velocities labeled in the 
lower right corner of each panel), contours are drawn at 0.21\,Jy beam$^{-1}$ 
(3\,$\sigma$) intervals, starting at 0.18\,Jy beam$^{-1}$.  A linear color 
scale represents the mean velocity of each channel, with red and blue denoting 
receding and approaching velocities, respectively.  \label{fig:co21}}
\end{figure}

Each individual dataset was calibrated independently with the {\tt MIR} 
software package.  The bandpass response was corrected based on observations of 
bright quasars, and broadband continuum channels were generated from the 
central portions of all line-free spectral chunks.  The visibility amplitude 
scale was derived by bootstrapping the gain calibrator (quasar) flux densities 
from the observations of Uranus, Callisto, or Ceres, with a systematic 
uncertainty estimated at 10-15\%.  The antenna-based complex gain response of 
the system was determined with reference to J1924-292, and the quality of the 
phase transfer was assessed using the observations of J1733-130.  That 
comparison suggests only a small amount of ``seeing" ($\sim$0\farcs1) was 
introduced by atmospheric phase noise (or small baseline errors), consistent 
with the excellent observing conditions.  After applying the appropriate (and 
small) phase shifts to account for the V4046 Sgr proper motion 
\citep[$\mu_{\alpha} \cos{\delta} = 0\farcs003$\,yr$^{-1}$, $\mu_{\delta} = 
-0\farcs052$\,yr$^{-1}$;][]{zacharias10} and confirming that the continuum 
amplitudes from different array configurations were consistent on overlapping 
baseline lengths, the visibility datasets from each observation were combined.  
The observations of the dust continuum and CO isotopologue emission will be 
presented in a separate article; the focus here will be solely on the $^{12}$CO 
$J$=2$-$1 emission.  Note that although the 2009 data were originally presented 
by \citet{rodriguez10}, those data have been re-calibrated here for 
consistency (and modest improvements).  

\begin{figure}[t!]
\epsscale{1.1}
\plottwo{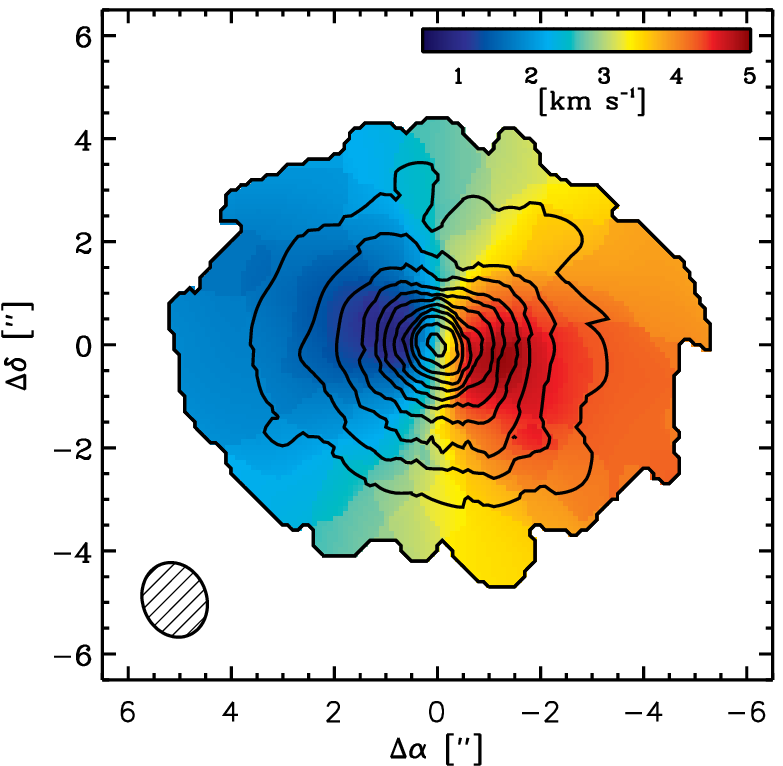}{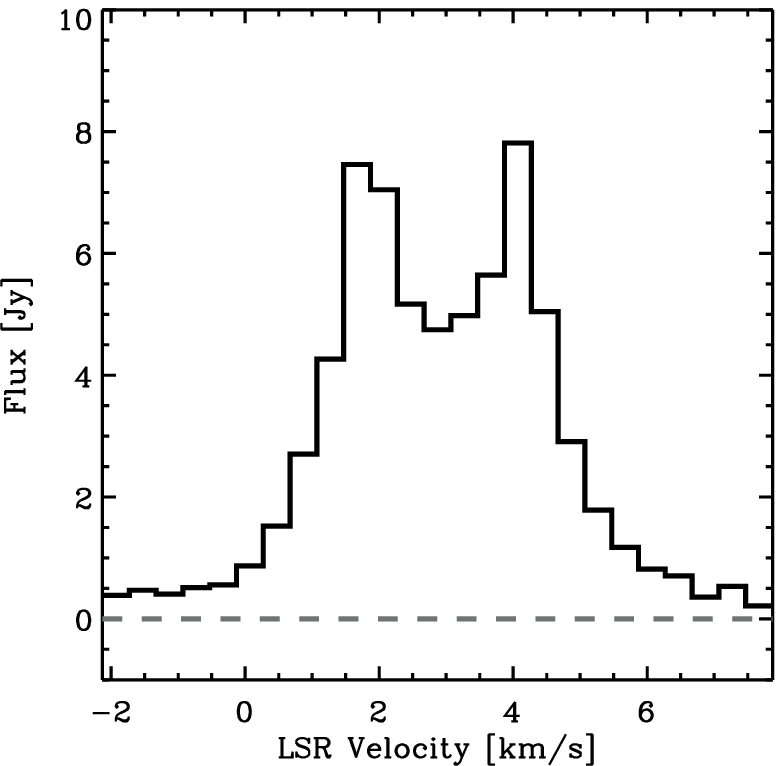}
\figcaption{({\it left}) The CO $J$=2$-$1 moment maps for the V4046 Sgr disk, 
on the same angular scale as in Figure \ref{fig:co21}.  The 0$^{th}$ moment 
(velocity-integrated intensity) map is overlaid in contours shown at 
5\,$\sigma$ intervals (0.35\,Jy km s$^{-1}$ beam$^{-1}$), starting at the 
3\,$\sigma$ level (0.21\,Jy km s$^{-1}$ beam$^{-1}$).  The $1^{st}$ moment 
(intensity-weighted velocities) map is shown in color, with a scale bar for 
reference.  ({\it right}) The integrated CO $J$=2$-$1 line profile from the 
V4046 Sgr disk, constructed from a square box 10\arcsec\ on a side centered at 
the stellar position, and assuming a 2\,$\sigma$ threshold for inclusion.  
\label{fig:mom}}
\end{figure}

The CO visibilities were continuum-subtracted and truncated outside a projected 
baseline length of 200\,k$\lambda$ to reduce the data volume and improve the 
signal-to-noise ratio.  They were then Fourier inverted, deconvolved with the 
{\tt CLEAN} algorithm, and restored with a synthesized beam using the {\tt 
MIRIAD} package.  The naturally-weighted spectral images shown as channel maps 
in Figure 1 were synthesized on a 0.4\,km s$^{-1}$ smoothed velocity scale with 
a $1\farcs55\times1\farcs29$ beam (at a position angle of 24\degr).  The 
typical RMS noise level in each channel is 70\,mJy beam$^{-1}$.  There is CO 
emission firmly detected ($>$3\,$\sigma$) out to $\pm$4.8\,km s$^{-1}$ from the 
systemic velocity, estimated to be $v({\rm LSR}) = +2.87\pm0.05$\,km s$^{-1}$ 
(corresponding to $-6.26\pm0.05$\,km s$^{-1}$ in the heliocentric frame), with 
an integrated intensity of $27.7\pm2.8$\,Jy km s$^{-1}$ and a peak flux density 
of $1.48\pm0.16$\,Jy beam$^{-1}$ ($17\pm2$\,K; a peak S/N = 21), including the 
calibration uncertainties.  Figure \ref{fig:mom} shows a map of the 
velocity-integrated CO intensities (0$^{\rm th}$ moment; {\it contours}) 
overlaid on the intensity-weighted velocities (1$^{\rm st}$ moment; {\it 
colorscale}), as well as a spatially integrated CO spectrum.  These data 
exhibit a molecular gas disk with a clear rotation pattern, from east 
(blueshifted) to west (redshifted), and suggest a modest inclination angle to 
the line of sight.  Near the systemic velocity, the CO emission subtends 
$\sim$5\arcsec\ in radius \citep[$\sim$365\,AU; see][]{rodriguez10}.

\section{Modeling Analysis} \label{sec:modeling}

The fundamental goal here is to derive a dynamical estimate of the central 
stellar mass based on the kinematic properties of the V4046 Sgr gas disk.  In 
order to extract $M_{\ast}$ from a measurement of the disk velocity field, we 
need to construct a detailed physical model of the disk structure.\footnote{For 
the sake of convenient and general notation, we will refer to the central 
stellar mass as $M_{\ast}$.  In this particular case, $M_{\ast}$ corresponds to 
the sum of the stellar masses in the V4046 Sgr binary.}  We adopt a modeling 
formalism motivated by \citet{beckwith93}, which makes three basic assumptions 
about disk properties.  First, the disk material is assumed to be orbiting in 
Keplerian rotation around a point mass, meaning the central stars are treated 
as one object that dominates the disk velocity field.  Previous work has found 
little evidence to support (simple parametric) deviations from Keplerian 
$v$-fields in disks \citep[e.g., see][]{simon00}; the corollary criterion that 
the disk mass is only a small fraction of the stellar mass ($M_d/M_{\ast} \ll 
1$) can be confirmed {\it a posteriori} (see \S 4).  Second, the disk is 
assumed to be geometrically thin at all radii.  Although \citet{pietu07} argued 
that this may not be valid at very large radii ($>$800\,AU), the higher 
$M_{\ast}$ for the V4046 Sgr binary (which decreases the disk scale height; see 
below) and the limited extent of the CO emission from its disk ($<$400\,AU) 
suggest it is a reasonable approximation in this case.  And third, in the 
context of the disk structure, we assume that the gas is vertically isothermal 
and in hydrostatic pressure equilibrium.  A more realistic model would include 
a temperature inversion \citep[e.g.,][]{dalessio98}.  However, excluding that 
kind of added complexity does not diminish the accuracy of an $M_{\ast}$ 
estimate: the emission from a single CO line is generated in a narrow vertical 
layer of the disk atmosphere, which has a roughly constant temperature 
\citep{dartois03}.  

The model for the gas density distribution is constructed in cylindrical 
coordinates ($r$, $\phi$, $z$),
\begin{equation}
\rho(r,z) = \frac{\Sigma}{\sqrt{2\pi} H_p} \exp \left[ -\frac{1}{2} \left( \frac{z}{H_p} \right)^2 \right],
\end{equation}
where $\Sigma$ is the radial surface density profile and $H_p$ is the pressure 
scale height at each radius.  We assume a parametric version of the former that 
is appropriate for an accretion disk with a static, power-law viscosity profile 
\citep{lynden-bell74,hartmann98} and is currently the basis for most dust-based 
disk density measurements \citep[e.g.,][]{andrews09,andrews10}, 
\begin{equation} 
\Sigma = \Sigma_c \left(\frac{r}{r_c}\right)^{-\gamma} \exp \left[-\left(\frac{r}{r_c}\right)^{2-\gamma}\right],
\end{equation}
where $\gamma$ is a density gradient, $r_c$ is a characteristic radius, and 
$\Sigma_c$ is a normalization equivalent to $e \cdot \Sigma(r_c)$.  The scale 
heights are calculated with the explicit assumption of hydrostatic equilibrium 
and a constant vertical temperature profile, such that
\begin{equation}
H_p = \frac{c_s}{\Omega} = \left( \frac{kT}{\mu m_{\rm H}} \cdot \frac{r^3}{G M_{\ast}} \right)^{1/2},
\end{equation}
where $c_s$ is the sound speed, $\Omega$ is the angular velocity, $T$ is the 
radial temperature profile, $k$ is the Boltzmann constant, $G$ is the 
gravitational constant, $m_{\rm H}$ is the mass of a hydrogen atom, and $\mu = 
2.37$ is the mean molecular weight of the gas.  We further assume a simple 
power-law behavior for the radial temperature profile,
\begin{equation}
T = T_{10} \left(\frac{r}{{\rm 10\,AU}}\right)^{-q},
\end{equation}
where $q$ is a temperature gradient and $T_{10}$ is the gas temperature at a 
radius of 10\,AU.  The parametric descriptions in Equations (1)-(4) completely 
characterize the temperature and density structure of a model gas disk.  The 
gas kinematics are described by a Keplerian velocity field, 
\begin{equation}
v_{\phi}(r) = v_k = \sqrt{\frac{G M_{\ast}}{r}} \, ; \,\,\,\, v_r = v_z = 0,
\end{equation}
meaning there is only rotation, and no net motion in the radial or vertical 
dimensions.  

Given a physical disk structure, we can construct a model of the CO $J$=2$-$1 
emission with the added assumption that the energy levels relevant for this 
transition are populated according to local thermodynamic equilibrium (LTE).  
Although the LTE approximation is not always valid in protoplanetary disks, 
\citet{pavlyuchenkov07} demonstrated that it is an appropriate simplification 
for the low energy and high optical depths associated with this particular 
species and transition.  We model the emission line intensity distribution by 
integrating the radiative transfer equation along each sight line $s$, so that
\begin{equation}
I_{\nu} = \int_0^{\infty} K_{\nu}(s) S_{\nu}(s) e^{-\tau_{\nu}(s)} ds,
\end{equation}
where $K_{\nu}(s)$ is the absorption coefficient, $S_{\nu}(s) = B_{\nu}(T)$ is 
the source function (here equivalent to the Planck function), and 
$\tau_{\nu}(s) = \int_0^s K_{\nu}(x) dx$ is the optical depth along the 
line-of-sight.  The absorption coefficient is the sum of contributions from 
dust and gas.  For the former, we assume $K_{\nu}(s)_{\rm dust} = \rho(s) \, 
\zeta \, \kappa_{\nu}$, where the dust-to-gas mass ratio is $\zeta = 0.01$ and 
the grain opacity is $\kappa_{\nu} \approx (\nu/$100\,GHz)\,cm$^2$ g$^{-1}$ 
\citep{beckwith90}.  In the particular case of the V4046 Sgr disk, the 
contribution of dust to the absorption coefficient is effectively negligible 
compared to the gas; the specific choices of $\zeta$ and $\kappa_{\nu}$ (within 
reasonable limitations) have no tangible effect on our results (moreover, 
continuum emission is removed from the data before our modeling analysis).  

The absorption coefficient of the CO gas is calculated from the transition 
cross section weighted by the population of the lower energy level, $\ell$, 
such that $K_{\nu}(s)_{\rm co} = n_{\ell}(s) \sigma_{\nu}(s)$ (note that in 
this specific case, $\ell = 1$).  Since we assume the line is thermally 
populated in LTE, the level populations are determined by the local disk 
temperature via the Boltzmann equation, 
\begin{equation}
n_{\ell}(s) = \frac{X_{\rm co} \, \rho(s)}{\mu \, m_{\rm H}} \cdot \frac{g_{\ell}}{Z} \, \exp \left( - \frac{E_{\ell}}{kT(s)} \right),
\end{equation}
where $E_{\ell}$ is the transition energy, $g_{\ell} = 2{\ell} + 1$ is the 
statistical weight, $Z$ is the partition function, and $X_{\rm co}$ is the CO 
fractional abundance (assumed to be constant everywhere in the disk).  The 
details of the emission line model are encoded in the absorption cross section,
\begin{equation}
\sigma_{\nu}(s) = \phi_{\nu}(s) \cdot \sigma_0 (1 - e^{-h \nu / k T(s)}),
\end{equation}
where $\phi_{\nu}$ is the line profile function and the integrated cross 
section is
\begin{equation}
\sigma_0 = \frac{h\nu}{4\pi} \cdot \frac{g_{\ell+1}}{g_{\ell}} B_{21} = \frac{c^2}{8 \pi \nu^2} \cdot \frac{g_{\ell+1}}{g_{\ell}} A_{21},
\end{equation}
where we have used the Einstein relation in the last equality to express the 
Einstein-$B$ coefficient in terms of the Einstein-$A$ coefficient provided by 
the LAMDA molecular database \citep{scholier05}.  The line profile function 
naturally determines the shape of the emission line, where a given frequency 
$\nu$ corresponds to an intrinsic Doppler velocity $v_d = (c/\nu_0)(\nu-\nu_0)$ 
relative to the line center, $\nu_0$.  The gas in a given disk model has a 
projected, line-of-sight velocity field given by $v_{\rm obs} \approx 
v_{\phi}(r) \cos{\phi} \sin{i_d}$, where $i_d$ is the inclination of the disk 
relative to the observer \citep[such that $i_d = 90$\degr\ is edge-on; for the 
geometry, see][]{isella07}.  The line profile shape at each frequency is 
determined by the difference between the Doppler and line-of-sight velocities, 
\begin{equation}
\phi_{\nu}(s) = \frac{c}{\sqrt{\pi} \nu_0 \Delta v} \exp \left[ - \left( \frac{v_d-v_{\rm obs}}{\Delta v}\right)^2 \right],
\end{equation}
and the effective line width, $\Delta v$.  The latter is comprised of the 
quadrature sum of contributions from thermal and non-thermal broadening terms,
\begin{equation}
\Delta v = \left( \frac{2 k T(s)}{m_{\rm co}} + \xi^2 \right)^{1/2},
\end{equation}
where $m_{\rm co}$ is the mean molecular weight of CO and $\xi$ is the 
contribution from microturbulence.  We assume that the turbulent velocity width 
$\xi$ is constant throughout the disk.

The \citet{beckwith93} model formalism highlighted above is admittedly 
complex.  A single model is completely described by 14 parameters: four 
quantify the distribution of CO densities \{$X_{\rm co}$, $\Sigma_c$, $R_c$, 
$\gamma$\}, two characterize the temperature structure (and therefore the 
vertical density structure) \{$T_{10}$, $q$\}, three play pivotal roles 
describing the disk kinematics \{$M_{\ast}$, $\xi$, $\nu_0$\}, and the 
remaining five relate the model to the observations -- i.e., convert from 
physical coordinates in the disk to the observed plane projected on the sky -- 
including the disk inclination and major axis position angle \{$i_d$, PA$_d$\}, 
distance \{$d$\}, and disk center \{$\alpha_0$, $\delta_0$\}.  Some of these 
parameters can be reliably determined in a simple way, and then fixed to 
facilitate the data modeling problem (with no tangible impact on the quality of 
the $M_{\ast}$ determination).  In this case, the systemic LSR velocity 
(effectively $\nu_0$) was measured directly from the SMA data and set to $v_0 = 
+2.87$\,km s$^{-1}$.  The disk center coordinates were estimated from both the 
dust continuum and CO emission morphology at the systemic velocity (see 
Fig.~\ref{fig:co21}) and set to $\alpha_0 = 18^{\rm h}14^{\rm m}10\fs48$, 
$\delta_0 = -32\degr47\arcmin35\farcs08$ (J2000), coincident with the composite 
V4046 Sgr stellar position in the UCAC3 catalog \citep{zacharias10}.  We 
adopted the distance inferred by \citet{torres06} from the kinematic parallax 
(moving cluster) technique, $d = 73$\,pc (but see \S 5 regarding alternative 
values).  

Some additional practical simplications can be made, since the focus here is on 
the velocity field and not the disk structure details.  Because the line 
opacity is so much greater than the dust opacity, the normalization parameters 
$X_{\rm co}$ and $\Sigma_c$ are not independent: we can only constrain their 
product.  In practice, we adopt a joint CO disk mass parameter, $M_{\rm co} = 
X_{\rm co} \Sigma_c (2 \pi r_c^2)/(2 - \gamma)$, for that purpose.  Moreover, 
after extensive experimentation, we concluded that two more parameters should 
be fixed.  First, we found that the disk orientation was very well-determined 
(within 1\degr) at PA$_d = 76\degr$ (E of N), so that continued iteration on a 
more precise value was a waste of computational resources: fixing this 
parameter has negligible quantitative impact on the other model parameters or 
their uncertainties.  And second, we fixed the surface density gradient, 
$\gamma = 1$.  The key parameters that set the density profile, \{$\gamma$, 
$r_c$\} are anti-correlated and strongly degenerate; statistically 
indistinguishable model fits can be found over a large range of these 
parameters.  The degeneracy is remarkably narrow, and could easily be missed 
with standard minimization algorithms (resulting in local $\chi^2$ minima with 
misleadingly tight constraints on the gradient).  Fortunately, this degeneracy 
has minimal impact on the parameters most relevant for characterizing the disk 
velocity field: the precision and accuracy of a $M_{\ast}$ determination are 
not notably affected by the uncertainties in \{$\gamma$, $r_c$\}.  With that in 
mind, we have made the modeling process more tractable with a fixed $\gamma$.

Making use of those simplifications, a synthetic CO spectral datacube can be 
calculated by specifying 7 free parameters, \{$M_{\rm co}$, $r_c$, $T_{10}$, 
$q$, $\xi$, $i_d$, $M_{\ast}$\}.  That model datacube is then resampled at the 
observed velocities and spatial frequencies and processed in the same way as 
the data (see \S 2) to produce a set of synthetic spectral visibilities.  Those 
model visibilities are evaluated with respect to the data by computing a 
composite $\chi^2$ value, summed over the real and imaginary components in 25 
spectral channels.  The best-fit set of model parameters and their associated 
uncertainties were determined with a Monte Carlo Markov Chain (MCMC) technique, 
using the affine invariant ensemble sampler developed by \citet[][see also 
Foreman-Mackey et al.~2012]{goodman10}, using a jump probability $\propto 
e^{-\Delta \chi^2/2}$.  To our knowledge, this is the first application of 
these more sophisticated MCMC methods for parameter estimation in this 
particular context: previous studies have relied on downhill simplex routines 
\citep[e.g.,][]{guilloteau98,simon00}, sometimes with clever modifications to 
address asymmetric uncertainties \citep{pietu07}.  Although comparatively the 
MCMC technique is computationally expensive, this Bayesian treatment has the 
distinct advantage of providing the posterior probability distributions for 
each parameter in the complex, multi-dimensional parameter-space of the 
underlying disk model.  

\begin{figure}[t!]
\epsscale{1.00}
\plotone{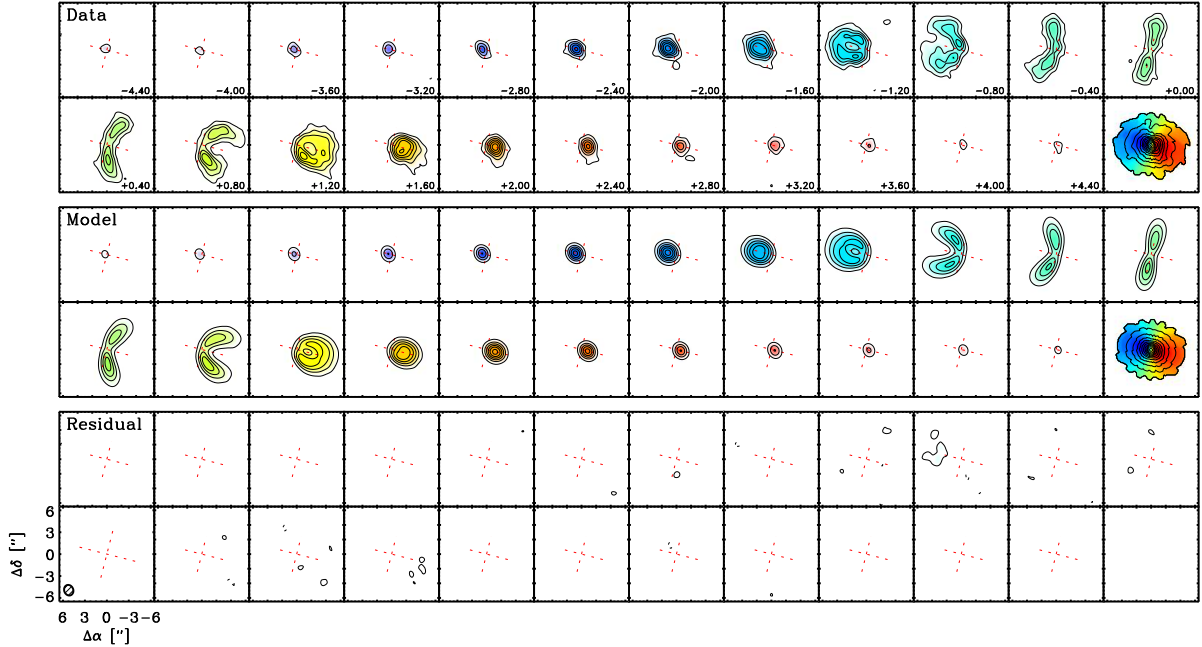}
\caption{Channel maps of the CO $J$=2$-$1 emission line from the SMA
observations ({\it top}), the best-fit disk model ({\it top}), and the imaged
residuals ({\it bottom}) with the channel velocity increasing from the top left
to the bottom right.  The last panel shows the 0$^{\rm th}$ ({\it contours})
and 1$^{\rm st}$ ({\it colorscale}) moment maps as in Figure \ref{fig:mom}.
The best-fit model parameters are listed in Table \ref{tab:params}.
\label{fig:bestfit}}
\end{figure}

\section{Results} \label{sec:results}

\begin{figure}[t!]
\epsscale{1.00}
\plotone{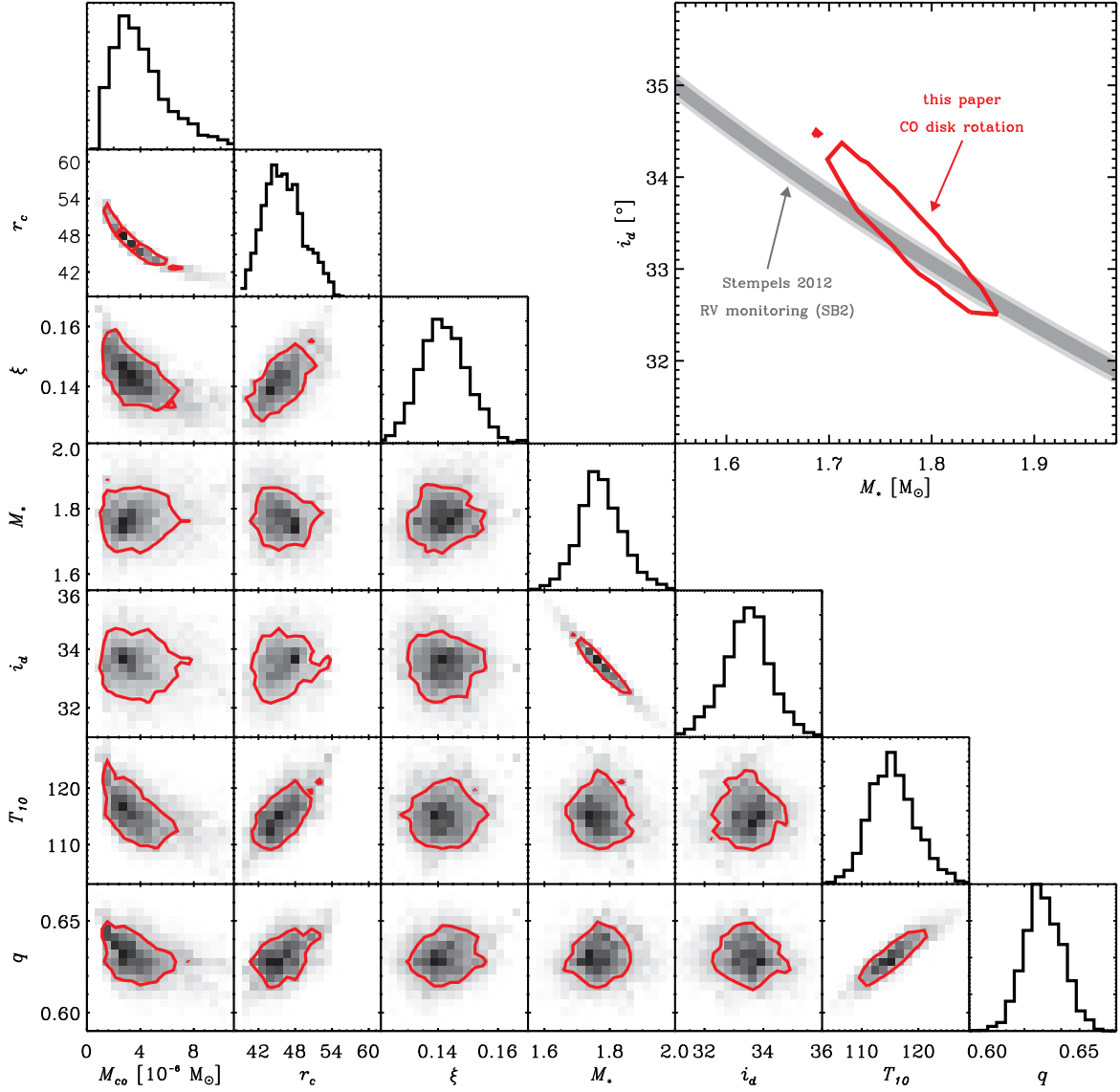}
\caption{Marginalized pararameters of the sampled posterior distrubution. The 68\% confidence contour is given by the red contour with each binned distribution shown by the linear greyscale.
\label{fig:posterior}}
\end{figure}

The best-fit model derived from the SMA data is compiled in Table 
\ref{tab:params}, which lists the mode (peak) of the posterior probability
distribution for each parameter along with its uncertainty, based on the range
of values that encompass 68\%\ of the distribution area (i.e., 1\,$\sigma$
uncertainties for a Gaussian distribution).  Figure \ref{fig:bestfit} makes a 
direct comparison between the data and the best-fit model in the spectral image 
plane, demonstrating the fit quality with only low-level (statistically 
insignificant) residuals present in the channel maps.  The best-fit model has a 
reduced $\tilde{\chi}^2 = 967,427/977,000 = 0.9902$.  A more detailed view of 
the multi-dimensional results of the modeling analysis is presented in Figure
\ref{fig:posterior}, where we have taken the 30,000 MCMC samples computed and 
marginalized over parameter subsets to display one and two-parameter posterior 
probability distributions.  The marginalized distributions for individual 
parameters are single-peaked, while the paired distributions highlight internal 
degeneracies in the model.  For example, there are tight correlations between 
normalizations and sizes (e.g., \{$M_{\rm co}$, $r_c$\}, to conserve the 
integrated line intensity; this would be modified were $\gamma$ a free 
parameter) or gradients \citep[e.g., \{$T_{10}$, $q$\}, to maintain an 
appropriate temperature in the outer disk;][]{mundy96,aw07}, as well as more 
subtle associations amongst parameters related to line broadening 
\citep[amongst $r_c$, $T_{10}$, and $\xi$;][]{guilloteau98,hughes11}.  The 
degeneracy of most interest here is the \{$M_{\ast}$, $i_d$\} anti-correlation, 
which is a natural consequence of reproducing the observed line-of-sight 
velocity pattern, $v_{\rm obs} \propto v_k \sin{i_d} \propto \sqrt{M_{\ast}} 
\sin{i_d}$ (i.e., $M_{\ast} \propto 1/\sin^2{i_d}$).

The best-fit model parameters that describe the V4046 Sgr disk structure are 
typical for young protoplanetary disks.  The adopted (fixed) surface density 
gradient ($\gamma = 1$) and characteristic radius ($r_c = 45^{+5}_{-3}$\,AU) 
lie near the median values for a survey of disk structures in the Ophiuchus 
region \citep{andrews09,andrews10}; the temperature profile ($T_{10} = 
115\pm5$\,K, $q = 0.63\pm0.01$; and therefore vertical density distribution) 
and turbulent linewidth ($\xi = 0.14\pm0.01$\,km s$^{-1}$) are comparable to 
those inferred from CO emission in other T Tauri disks 
\citep[e.g.,][]{guilloteau98,pietu07,hughes08,hughes11}.  If we assume the CO 
mass fraction found in dark clouds ($X_{\rm co} \approx 10^{-4}$) is also 
applicable in the disk, the molecular density normalization parameter $M_{\rm 
co}$ implies a modest gas mass, $2.8^{+3.7}_{-1.5} \times 
10^{-2}$\,M$_{\odot}$, validating {\it a posteriori} our assumption that 
self-gravity is negligible ($M_d/M_{\ast} \approx 10^{-2} \ll 1$).  In any 
case, our simple treatment of the disk structure is really only used as a means 
to an end: the focus here is to place a firm constraint on the key parameter 
that determines the behavior of the disk velocity field -- the central stellar 
mass, $M_{\ast}$.

We infer a total stellar mass of $M_{\ast} = 1.75^{+0.09}_{-0.06}$\,M$_{\odot}$ 
for the V4046 Sgr spectroscopic binary, a formally precise constraint with a 
relative uncertainty of 3-5\%.  Figure \ref{fig:posterior} demonstrates that 
this estimate of $M_{\ast}$ is essentially independent of the disk structure 
parameters in the model.  The sole relevant degeneracy is with the disk 
inclination, which we infer to be $i_d = 33\fdg5^{+0.7}_{-1.4}$ (a relative 
precision of $\sim$2-5\%).  Within the quoted uncertainties, these 
\{$M_{\ast}$, $i_d$\} values are consistent with the previous determinations by 
\citet{rodriguez10}, who used a simple $\chi^2$ grid search and an altogether 
different underlying disk structure model.  These \{$M_{\ast}$, $i_d$\} 
measurements based on the circumstellar gas disk kinematics can be directly 
compared with the joint (degenerate) constraints on \{$M_{\ast}$, $i_{\ast}$\} 
imposed by RV measurements of the spectroscopic binary itself.  Although 
various groups have studied this SB2 system 
\citep{quast00,stempels04,donati11}, by far the most robust constraints on the 
stellar parameters come from the long-term, extensive, and high resolution 
optical spectroscopic monitoring campaign conducted by \citet{stempels12}.  In 
that work, the RV data are found to be best explained with a binary that has a 
total mass $M_{\ast} (\sin{i_{\ast}})^3 = 0.2923\pm0.0007$\,M$_{\odot}$.  The 
inset in Figure \ref{fig:posterior} confirms that our disk kinematics 
constraints and the RV constraints by \citet{stempels12} coincide in the 
stellar mass$-$inclination plane: the two {\it completely independent} methods 
find the same results well within their formal 1\,$\sigma$ uncertainties.  
Adopting our best estimate of $M_{\ast}$ and propagating the relevant 
uncertainties, the RV constraints suggest that $i_{\ast} = 33\fdg42\pm0.01$.  
Therefore, we quantitatively confirm the finding of \citeauthor{rodriguez10}: 
the V4046 Sgr spectroscopic binary and its associated, large-scale circumbinary 
disk are co-planar within $|i_d - i_{\ast}| \approx 0.1\pm1\degr$ across 
$\sim$4 orders of magnitude in radial scale.

\clearpage

\section{Discussion}\label{sec:discuss}

We have presented spatially and spectrally resolved Submillimeter Array 
observations of the $^{12}$CO $J$=2$-$1 line emission from the isolated, 
nearby, and gas-rich circumstellar disk that orbits the close, pre-main 
sequence double-lined spectroscopic binary V4046 Sgr.  Adopting a simple 
parametric model for the structure of a flared Keplerian disk, we employ a 
Monte Carlo Markov Chain technique to infer the properties of the disk velocity 
field from these data, and thereby provide a robust statistical estimate for 
the total mass of the central binary.  We find that these CO line data are best 
reproduced for a binary mass $M_{\ast} = 1.75^{+0.09}_{-0.06}$\,M$_{\odot}$ and 
a disk viewing angle inclined by $i_d = 33\fdg5^{+0.7}_{-1.4}$ from face-on.  
Those values are in excellent agreement with the completely independent 
inferences of \citet{stempels12}, made from their extensive radial velocity 
monitoring campaign of the spectroscopic binary itself.  The orbital planes of 
the binary ($a = 0.045$\,AU) and its associated circumbinary disk (on radial 
scales out to $\sim$400\,AU) are aligned within $\sim$0.1$\pm$1\degr.  These 
results demonstrate that, despite its complexity, the disk-based kinematic 
method for estimating the masses of young stars is both {\it precise} (at the 
level of a few percent) and {\it accurate} in an absolute sense, as verified 
here by an entirely independent dynamical constraint.  

The disk-based dynamical estimate of $M_{\ast}$ can be combined with the RV 
constraints of \citet{stempels12} and other information from the literature to 
assess the predictions of pre-MS stellar evolution models.  Coupling our 
$M_{\ast}$ estimate with the stellar mass ratio ($q$) and mass function 
($M_{\ast} (\sin{i_{\ast}})^3$) determined by \citeauthor{stempels12}, we can 
derive the masses of the individual stellar components in the V4046 Sgr binary, 
$0.90\pm0.05$ and $0.85\pm0.04$\,M$_{\odot}$, and their orbital inclination, 
$i_{\ast} = 33\fdg42\pm0.01$.\footnote{Here and throughout, we conservatively 
adopt the larger of all asymmetric uncertainties for simplicity.}  Assuming 
that the stellar rotation axes are aligned with the binary orbital axis 
\citep[a spin-orbit alignment presumably induced by tides in this close, 
circularized system; e.g.,][]{zahn77,hut81,melo01}, the radial and {\it 
rotational} velocities measured for each stellar component by 
\citeauthor{stempels12} can be used to determine the stellar radii, 
$1.25\pm0.04$ and $1.21\pm0.04$\,R$_{\odot}$.  \citet{stempels04} inferred 
spectral types of K5 and K7 in this system, corresponding to effective 
temperatures of $4350\pm240$ and $4060\pm210$\,K for the standard conversion 
advocated by \citet{schmidt-kaler82}; an ambiguity of one spectral subclass has been 
assumed to estimate the temperature uncertainties.  Those measurements imply 
that the individual stellar luminosities are $0.50\pm0.11$ and 
$0.36\pm0.08$\,L$_{\odot}$ (a total luminosity of $0.86\pm0.14$\,L$_{\odot}$). 

\begin{figure}[t!]
\epsscale{0.6}
\plotone{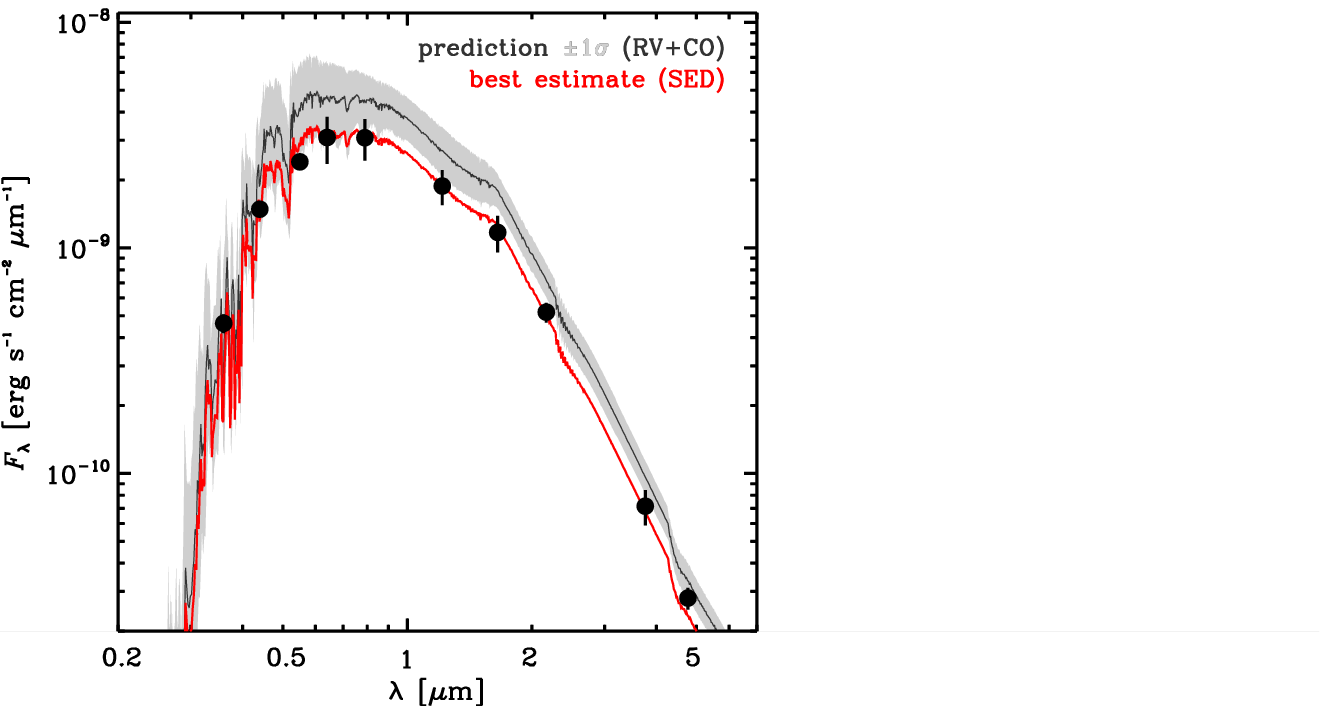}
\figcaption{The composite optical/near-infrared broadband spectrum observed for
the V4046 Sgr binary (see text for references and discussion of uncertainties),
along with a model spectrum predicted from a combination of the CO rotation
curve analysis in \S 4 and the optical spectroscopic monitoring campaign of
\citet{stempels12} ({\it black} curve, with uncertainties shaded in
{\it gray}).  The {\it red} curve corresonds to that prediction scaled down by
$\sim$30\%\ in luminosity, to best reproduce the observations.  The data and
the predictions are marginally discrepant (1.4\,$\sigma$) in terms of their
normalization; potential causes of that minor difference are discussed in the
text.  \label{fig:spectrum} }
\end{figure}

Figure \ref{fig:spectrum} shows the composite optical/near-infrared spectrum 
expected from the V4046 Sgr binary, given the effective temperatures, stellar 
radii, and masses derived above from the combination of RV data and our 
disk-based measurements ({\it black curve}, with uncertainty in {\it shaded 
gray}).  This spectral energy distribution (SED) prediction was generated from 
an interpolated grid of \citet{lejeune97} model spectra, assuming the 
appropriate stellar gravities ($\log{g} = 4.20\pm0.02$ in both cases), 
negligible interstellar reddening ($A_V = 0$), and our adopted $d = 73$\,pc.  
Representative broadband photometric data are marked as the datapoints in 
Figure \ref{fig:spectrum}, constructed from the weighted mean magnitudes of 
\citet{hutchinson90}, \citet{strassmeier93}, \citet{jensen97}, and the 2MASS 
\citep{skrutskie06} and DENIS \citep{epchtein97} surveys: uncertainties were taken 
as the quadrature sum of the standard deviation of the weighted mean and the 
maximal deviation from the weighted mean, in an effort to more faithfully 
represent potential uncertainties due to variability.  A quick examination of 
Figure \ref{fig:spectrum} demonstrates that the spectral morphology of the data 
and model prediction are a good match, although the normalizations appear 
discrepant.  Scaling the (best-fit) predicted spectrum down by 
$\sim$30$\pm15$\%\ provides a good match to the observed spectrum; a 
composite luminosity of $0.60\pm0.13$\,L$_{\odot}$ is more appropriate ({\it 
red curve}).  Therefore, the predicted and observed composite binary 
luminosities are marginally ($\sim$1.4\,$\sigma$) different.  

Although formally the conflict in these luminosity values is not statistically 
significant, it is still interesting to consider some potential paths for a 
better reconciliation of all the data.  Perhaps the most straightforward means 
of doing so is to modify the assumed distance to the system: taken at face 
value, a $\sim$30\%\ increase to $d \approx 95$\,pc implies an inherently more 
luminous pair of stars with a spectrum that would be in good agreement with the 
observations.  However, that same re-normalization impacts the total stellar 
mass inferred from the CO data, with a linear scaling to $M_{\ast} \approx 
2.3$\,M$_{\odot}$ that introduces a substantial ($\sim$2\,$\sigma$) discrepancy 
in the \{$M_{\ast}$, $i$\}-plane between the disk-kinematics and RV techniques 
for estimating stellar parameters.  Although it hardly seems worthwhile to 
trade one marginal discrepancy for another (which is technically less 
marginal), there is no {\it a priori} reason that the orbital planes of the 
disk and binary need to be aligned with the precision inferred in \S 4.  
Unfortunately, little guidance is provided in the form of an uncertainty on the 
kinematic parallax measurement of \citet{torres06}.  In any case, discrepancies 
in effective temperatures and luminosities are not necessarily uncommon for 
active young stars \citep{stassun12}, and particularly for those in close 
binary systems \citep[e.g.,][]{gomezmaqueochew12}.

Ultimately, dwelling on a marginal luminosity discrepancy is not well 
justified, especially given the independent distance estimate from the 
\citet{torres06} study.  In the following, we adopt the stellar parameters 
inferred from the joint constraints of the CO disk spatio-kinematics and the 
optical RV studies of \citeauthor{stempels12} assuming $d = 73$\,pc, but use 
individual stellar luminosities scaled down by 30\%\ to best match the observed 
optical and near-infrared spectrum: $0.35\pm0.10$\,L$_{\odot}$ and $0.25\pm0.08$\,L$_{\odot}$ \citep[note that each component is slightly less luminous than 
reported by][]{donati11}.  With those results, we can make comparisons with the 
predictions of pre-MS evolution models to estimate the ages ($t_{\ast}$) and 
masses of the individual stars in the V4046 Sgr binary.  To accomplish that, we 
follow the Bayesian methodology of \citet{jorgensen05}, which employs a 
finely-interpolated grid of pre-MS models in the HR diagram to estimate the 
probability distributions of stellar mass and age given the measured values 
(and uncertainties) of effective temperatures and luminosities \citep[see 
also][]{gennaro12}.  If we define the observables as \{$x$, $y$\} = 
\{$\log{T}$, $\log{L}$\} with associated uncertainties \{$\sigma_x$, 
$\sigma_y$\}, and the pre-MS model predictions \{$\hat{x}(M_{\ast}$, 
$t_{\ast})$, $\hat{y}(M_{\ast}$, $t_{\ast})$\}, we can write the likelihood 
function as a multivariate Gaussian,
\begin{equation}
\mathcal{L}(\hat{x},\hat{y}|x, y) = \frac{1}{2\pi \sigma_x \sigma_y }  \exp \left\{ -\frac{1}{2} \left[ \frac{(x - \hat{x})^2}{\sigma_x^2} +  \frac{(y- \hat{y})^2}{\sigma_y^2} \right]\right\}.
\end{equation}
The best-fit \{$M_{\ast}$, $t_{\ast}$\} are then directly determined from the 
\{$\hat{x}$, $\hat{y}$\} that maximize the likelihood, with uncertainties that 
can be calculated from the shape of the likelihood distribution.  This 
procedure was conducted for four different pre-MS evolution models, assuming 
solar composition, a fractional deuterium abundance of $2\times10^{-5}$, and a 
convective mixing parameter $\alpha \approx 1.7$-2.0: \citet{dantona98}, 
\citet{baraffe98}, \citet{siess00}, and \citet{tognelli11}.

\begin{figure}[t!]
\epsscale{1.0}
\plotone{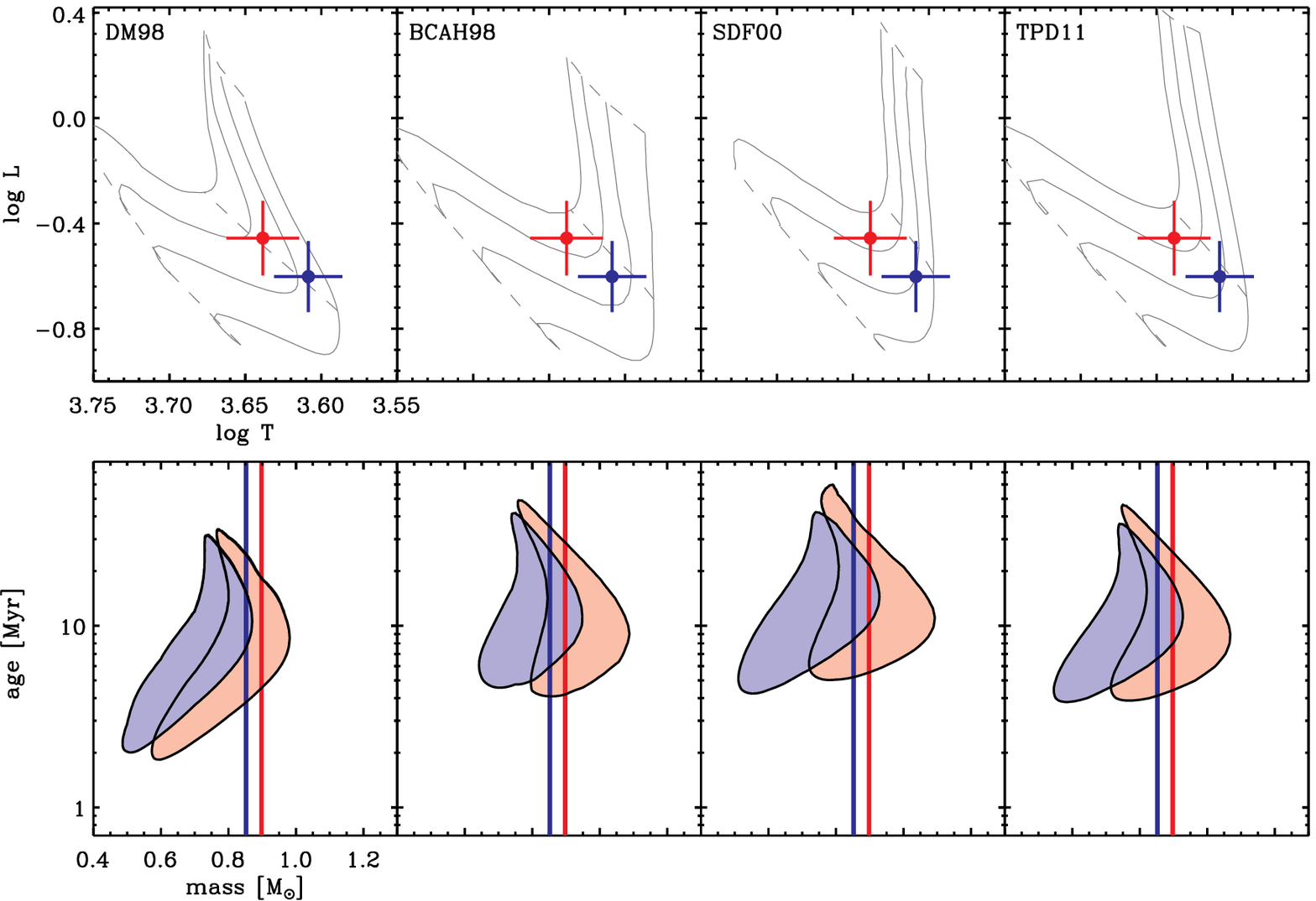}
\caption{({\it top}) The V4046 Sgr stellar properties (primary in {\it red}, 
secondary in {\it blue}) compared with the pre-MS evolutionary models of ({\it 
from left to right}) \citeauthor{dantona98} (DM98), \citeauthor{baraffe98} 
(BCAH98), \citeauthor{siess00} (SDF00), and \citet{tognelli11} (TPD11) in the 
HR diagram.  In each case, the mass tracks for $M_\ast = \{0.7, 0.8, 0.9, 
1.0$\}\,M$_\odot$ are shown as gray solid curves, with isochrones for $t_{\ast} 
= \{1, 10, 100\}$\,Myr denoted as gray dashed curves.  ({\it bottom}) The 68\% 
confidence intervals for the posterior likelihood distribution in the 
\{$M_{\ast}$, $t_{\ast}$\}-plane derived from the HR diagram for each model 
type, using the Bayesian methodology of \citet{jorgensen05}.  The vertical 
lines indicate the individual stellar masses as determined by combining the 
total mass from the disk analysis in \S 4 and the mass ratio derived from the 
spectroscopic analysis of \citet{stempels12}. \label{fig:HR}} 
\end{figure}

\begin{figure}[t!]
\epsscale{1.10}
\plottwo{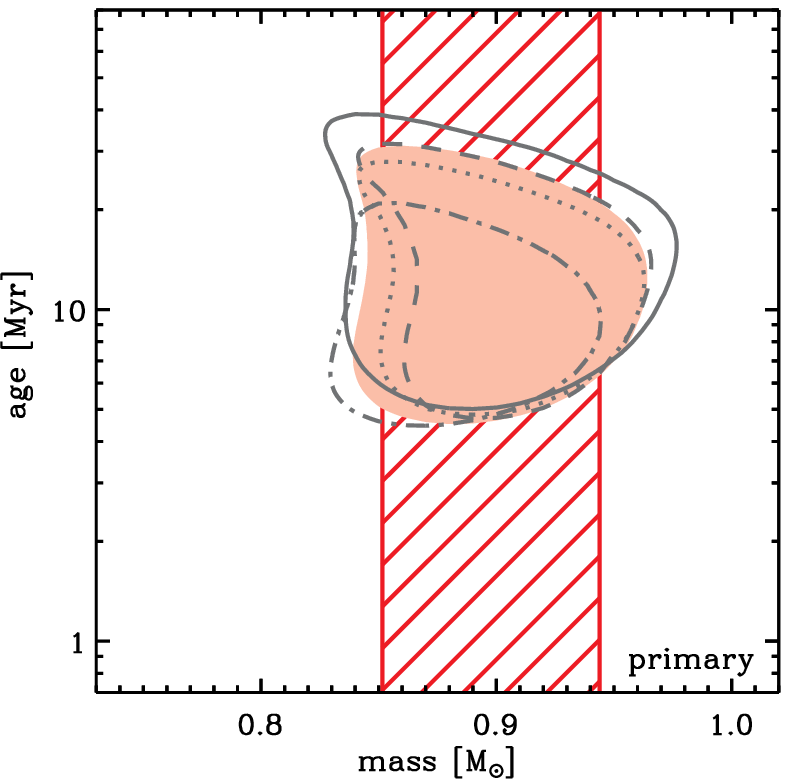}{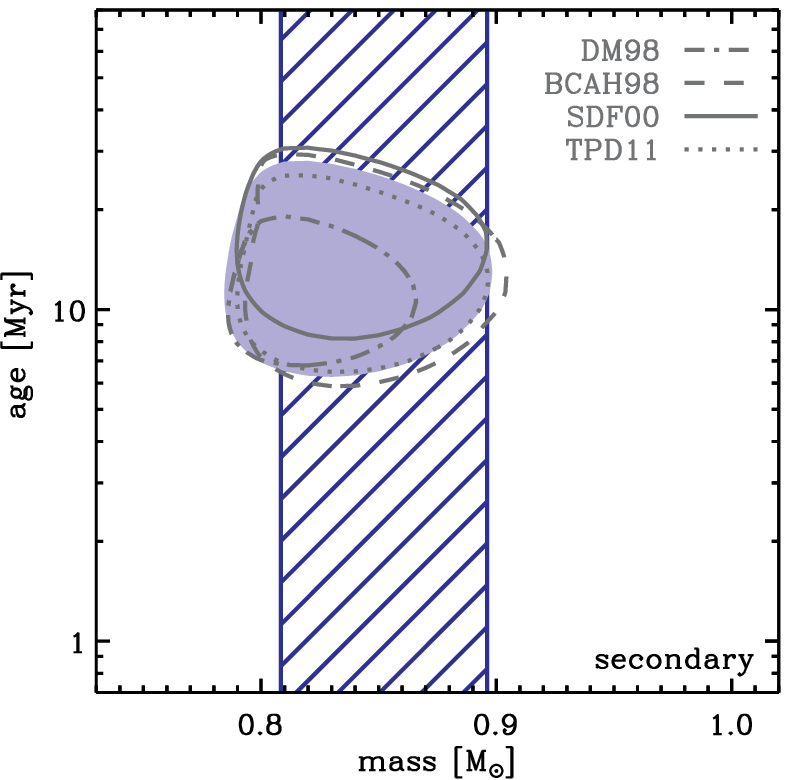}
\figcaption{The 68\% confidence intervals for the posterior likelihood 
distribution in the \{$M_{\ast}$, $t_{\ast}$\}-plane derived from the HR 
diagram for each model type ({\it see key in right panel}) overlaid together on 
the individual V4046 Sgr components ({\it left} primary; {\it right} 
secondary), now assuming the dynamical masses determined in \S 4 as Gaussian 
priors (the `+dyn' models references in Table \ref{tab:HR}).  The filled shaded 
region corresponds to the averaged likelihoods from the four different pre-MS 
models.  The dynamical masses and their uncertainties are marked as a hatched, 
shaded vertical column.  \label{fig:ages}}
\end{figure}

The results of these comparisons are presented in Table \ref{tab:HR}.  Figure 
\ref{fig:HR} shows the V4046 Sgr stellar properties that were inferred from 
each of the four reference sets of pre-MS evolutionary model tracks in the HR 
diagram.  All of the evolutionary models make predictions for the V4046 Sgr 
stellar masses that are remarkably consistent with each other and the dynamical 
masses inferred in \S 4.  The stellar masses that formally maximize the 
likelihood in Equation 12 are systematically $\sim$2-10\%\ below the best-fit 
dynamical masses, an under-prediction typical of pre-MS models regardless of 
the dynamical method used to estimate $M_{\ast}$ 
\citep{hillenbrand04,gennaro12}.  However, those modest discrepancies are not 
statistically significant, given the uncertainties on \{$L$, $T$\} and the 
dynamical masses.  An examination of Figure \ref{fig:HR} demonstrates that the 
V4046 Sgr stars are nearly evolved off the Hayashi track, having presumably 
developed radiative cores as expected for their solar-like masses.  The clear 
implication of their location in the HR diagram is that the V4046 Sgr binary is 
comparatively {\it old} for a pre-MS system, as has been previously reported by \citet{donati11} and \citet{kastner11}.  The models used here suggest a 
large range of ages are plausible, from $\sim$5-30\,Myr, and confirm that the 
binary components are coeval.  If we include a Gaussian prior representing the 
dynamical masses into the Bayesian analysis of the HR diagrams described above 
(labeled as `+dyn' in Table \ref{tab:HR}), we can infer a smaller range of 
acceptable ``dynamical" ages from each set of pre-MS models.  The different 
pre-MS model predictions with these dynamical priors are shown together, and 
averaged \citep[weighted by the posterior probability for each 
model;][]{hoeting99}, in Figure \ref{fig:ages}: the averaged results suggest 
ages of $12^{+17}_{-3}$ and $13^{+11}_{-3}$\,Myr for V4046 Sgr A and B, 
respectively.  The corresponding coeval age is 13$_{-3}^{+8}$\,Myr, in good 
agreement with the age constraints from the putative far-flung companion(s) 
V4046 Sgr C[D] \citep[at a separation of $\sim$12,350\,AU, or 2\farcm8 on the 
sky;][]{kastner11}.

\citet{torres06} suggested that the observed space motion of V4046 Sgr is 
consistent with a high probability of membership in the $\beta$ Pic moving 
group, a widespread and local ($d \approx 34\pm21$\,pc) stellar association 
with an age range estimated to be $\sim$8-20\,Myr 
\citep[e.g.,][]{zuckerman01,zuckerman04,torres08}.  The kinematic parallax 
calculated by Torres et al. ($d \approx 73$\,pc) and component ages estimated 
here for the V4046 Sgr binary are certainly supportive of that conclusion.  If 
that is the case, it is worth noting the uniqueness of the V4046 Sgr 
circumstellar environment: there are no other $\beta$ Pic moving group members 
that are known to host such a massive, long-lived, gas-rich accretion disk.  
Regardless of its original association, the proximity and advanced age of V4046 
Sgr make for a remarkable case study in both the long-term evolution of 
protoplanetary disk structure and the fundamental properties of pre-MS binary 
stars.

Although V4046 Sgr is a particularly striking example of the methodology behind 
disk-based dynamical estimates of stellar masses, we anticipate that these 
techniques will find considerably more use in the near future as the Atacama 
Large Millimeter Array (ALMA) project is completed.  Even with vastly improved 
data quality, this simplified model to extract the gas disk rotation curve from 
such interferometric observations (see \S 3) remains complex.  However, we have 
demonstrated that the method is robust, accurate, and precise, by using 
independent dynamical constraints from the V4046 Sgr spectroscopic binary 
radial velocity monitoring results.  In practice, these constraints and others 
like them \citep[notably for UZ Tau E; see][]{simon00,prato02} effectively 
serve as a check on the absolute calibration of the modeling technique 
described in \S 3.  The consistency between the independent dynamical 
constraints on the V4046 Sgr stellar masses validates the application of this 
procedure to dynamically ``weigh" single, isolated pre-MS stars based on their 
gas disk kinematics.  Along with an investment in accurate stellar luminosity 
and temperature measurements, ALMA observations of the molecular gas kinematics 
in circumstellar disks will usher in a new era of precision in the fundamental 
astrophysical properties of young stars.

\section{Summary} \label{sec:summary}

We have presented sensitive, high-resolution SMA observations of the $^{12}$CO 
$J$=2$-$1 line emission from the massive, gas-rich disk orbiting the 
double-lined spectroscopic binary star V4046 Sgr.  Using simple radiation 
transfer calculations for a disk structure model with a Keplerian velocity 
field, we fit the observed spectral visibilities using a stochastic model 
optimization technique that simultaneously infers model parameter values and 
their uncertanties.  Our specific focus has been on the key model parameters 
that describe the disk velocity field, with a goal of placing a firm, dynamical 
constraint on the mass of the central binary.  The key conclusions of our 
analysis are:
\begin{enumerate}
\item From modeling the CO line emitted by the circumbinary disk, we infer that 
the total stellar mass of the V4046 Sgr binary is $M_\ast = 
1.75_{-0.06}^{+0.09} M_\odot$, assuming the kinematic parallax distance of 
73\,pc estimated by \citet{torres06}.  That measurement is in excellent 
agreement with the independent dynamical constraints imposed from the analysis 
of an optical spectroscopic monitoring campaign of the radial velocity 
variations from the binary itself \citep{stempels12}.  

\item The mutual consistency of these distinct dynamical constraints on the 
stellar masses verifies that the mass determination based on the velocity field 
of the disk gas is accurate in an absolute sense, as well as remarkably precise 
(with a 3-5\% formal uncertainty on $M_{\ast}$).  

\item That same combination of constraints from millimeter and optical 
spectroscopic measurements confirm that the orbital planes of the stars and 
their accompanying circumbinary disk are co-planar, with inferred inclination 
angles ($i_d = 33\fdg5^{+0.7}_{-1.4}$ and $i_{\ast} = 33\fdg4\pm0.01$) that 
differ by $0.1\pm1$\degr\ over roughly four decades in radius, from $\sim$0.04 
to 400\,AU.

\item The inferred component masses of the V4046 Sgr binary are in uniformly 
good agreement with a variety of pre-MS evolution model predictions.  When 
combined with our dynamical constraints, those same models confirm the 
coevality of the binary components, and suggest an average age for the system 
of $13_{-3}^{+8}$\,Myr.  Therefore, V4046 Sgr hosts one of the oldest and 
nearest gas-rich primordial accretion disks currently known.  
\end{enumerate}

\acknowledgments We are very grateful to Joel Kastner, Guillermo Torres and Zachory Berta for 
some insightful discussions and suggestions, as well as to Germano Quast for 
being kind enough to provide additional information on their optical 
observations of the V4046 Sgr binary.  The SMA is a joint project between the 
Smithsonian Astrophysical Observatory and the Academia Sinica Institute of 
Astronomy and Astrophysics and is funded by the Smithsonian Institution and the 
Academia Sinica.

\begin{deluxetable}{ccr}
\tablecolumns{3}
\tablewidth{0pc}
\tablecaption{Disk Model Parameters \label{tab:params}}
\tablehead{\colhead{parameter} & \colhead{units} & \colhead{estimate}}
\startdata
$M_{\rm co}$&$ [10^{-6}~ M_\odot]$ & $2.8^{+3.7}_{-1.5}$ \\
$r_c$ &[AU] &$45_{-3}^{+5}$ \\
$q$ & \nodata &  $0.63\pm0.01$\\
$T_{10}$ &[K] & $115\pm5$\\
$\xi$& [km s$^{-1}$] & $0.14\pm0.01$ \\
${i_d}$ &[$^\circ$] &$33.5_{-1.4}^{+0.7}$ \\
$M_\ast$& [M$_\odot$] & $1.75_{-.06}^{+.09}$
\enddata
\tablecomments{See \S 3 for a description of the model parameters.  The 
uncertainties for each parameter are determined from the width of the posterior 
distribution that encapsulates 68\% of the models around the peak (equivalent 
to $1 \sigma$ for Gaussian random variables.}
\end{deluxetable}

\begin{deluxetable}{lcc|ccccc|cc}
\tablecolumns{10}
\tablewidth{0pc}
\tablecaption{Comparison with pre-MS Evolution Models\label{tab:HR}}
\tablehead{
\colhead{} & \multicolumn{4}{c}{V4046 Sgr A} & \colhead{} & \multicolumn{4}{c}{V4046 Sgr B} \\
\cline{2-5} \cline{7-10}
\colhead{} & \multicolumn{2}{c}{$M_{\ast}$ [M$_{\odot}$]} & \multicolumn{2}{c}{$t_{\ast}$ [Myr]} & \colhead{} & \multicolumn{2}{c}{$M_{\ast}$} & \multicolumn{2}{c}{$t_{\ast}$} \\
\cline{2-3} \cline{4-5} \cline{7-8} \cline{9-10}
\colhead{Models} & \colhead{HR} & \colhead{+dyn} & \colhead{HR} & \colhead{+dyn} &\colhead{} & \colhead{HR} & \colhead{+dyn} & \colhead{HR} & \colhead{+dyn}
}
\startdata
DM98   & $0.84^{+0.06}_{-0.10}$ & $0.88^{+0.03}_{-0.05}$ & $6^{+27}_{-1}$ & $9^{+10}_{-2}$ & & $0.76^{+0.06}_{-0.09}$ & $0.82^{+0.03}_{-0.05}$ & $8^{+19}_{-1}$ & $11^{+8}_{-3}$ \\
BCAH98 & $0.89^{+0.09}_{-0.12}$ & $0.90^{+0.04}_{-0.06}$ & $10^{+37}_{-1}$ & $18^{+10}_{-9}$ & & $0.80^{+0.07}_{-0.09}$ & $0.84^{+0.04}_{-0.05}$ & $10^{+14}_{-1}$ & $13^{+12}_{-10}$ \\
SDF00  & $0.83^{+0.13}_{-0.06}$ & $0.88^{+0.05}_{-0.05}$ & $12^{+52}_{-1}$ & $20^{+31}_{-8}$ & & $0.75^{+0.09}_{-0.10}$ & $0.83^{+0.04}_{-0.05}$ & $11^{+39}_{-2}$ & $16^{+13}_{-4}$ \\
TPD11  & $0.86^{+0.09}_{-0.11}$ & $0.89^{+0.04}_{-0.05}$ & $9^{+35}_{-1}$ & $14^{+10}_{-5}$ & & $0.78^{+0.07}_{-0.11}$ & $0.85^{+0.04}_{-0.04}$ & $9^{+21}_{-1}$ & $12^{+10}_{-3}$ \\
\hline
$\langle$mean$\rangle$ & $0.84_{-0.08}^{+0.11}$ & $0.89_{-0.05}^{+0.04}$ & $9_{-1}^{+43}$ & $12_{-3}^{+17}$ & & $0.77_{-0.11}^{+0.08}$ & $0.83_{-0.05}^{+0.04}$ & $10_{-3}^{+10}$ & $13_{-3}^{+11}$
\enddata
\tablecomments{Stellar mass and age estimates from the four different pre-MS 
evolution models are calculated using the HR observables \{$\log{L}$, 
$\log{T}$\}, following the methods outlined in \S 5.  We report results for 
both a uniform (HR) and Gaussian (+dyn) prior on the mass, with the latter 
based on the dynamical constraints.  The mean values are computed for all the 
models, following the technique outlined by \citet{hoeting99}.}
\end{deluxetable}


\clearpage

\begin{thebibliography}{}
\bibitem[Alibert et al.(2011)]{alibert11} Alibert, Y., Mordasini, C., \& Benz, W. 2011, \aap, 526, 63
\bibitem[Andrews \& Williams(2007)]{aw07} Andrews, S. M., \& Williams, J. P. 2007, \apj, 659, 705
\bibitem[Andrews et al.(2009)]{andrews09} Andrews, S. M., Wilner, D. J., Hughes, A. M., Qi, C., \& Dullemond, C. P. 2009, \apj, 700, 1502
\bibitem[Andrews et al.(2010)]{andrews10} Andrews, S. M., Wilner, D. J., Hughes, A. M., Qi, C., \& Dullemond, C. P. 2010, \apj, 723, 1241
\bibitem[Artymowicz \& Lubow(1994)]{artymowicz94} Artymowicz, P., \& Lubow, S. H. 1994, \apj, 421, 651
\bibitem[Baraffe et al.(1998)]{baraffe98} Baraffe, I., Chabrier, G., Allard, F., \& Hauschildt, P.~H.\ 1998, \aap, 337, 403 
\bibitem[Baraffe et al.(2002)]{baraffe02} Baraffe, I., Chabrier, G., Allard, F., \& Hauschildt, P. H. 2002, \aap, 382, 563
\bibitem[Baraffe et al.(2009)]{baraffe09} Baraffe, I., Chabrier, G., \& Gallardo, J. 2009, \apj, 702, L27
\bibitem[Baraffe \& Chabrier(2010)]{baraffe10} Baraffe, I., \& Chabrier, G. 2010, \aap, 521, 44
\bibitem[Bastian et al.(2010)]{bastian10} Bastian, N., Covey, K. R., \& Meyer, M. R. 2010, \araa, 48, 339
\bibitem[Beckwith et al.(1990)]{beckwith90} Beckwith, S. V. W., Sargent, A. I., Chini, R. S., \& G{\"{u}}sten, R. 1990, \aj, 99, 924
\bibitem[Beckwith \& Sargent(1993)]{beckwith93} Beckwith, S. V. W., \& Sargent, A. I. 1993, \apj, 402, 280
\bibitem[Boden et al.(2005)]{boden05} Boden, A.~F., Sargent, 
A.~I., Akeson, R.~L., et al.\ 2005, \apj, 635, 442 
\bibitem[Boden et al.(2007)]{boden07} Boden, A.~F., Torres, G., 
Sargent, A.~I., et al.\ 2007, \apj, 670, 1214 
\bibitem[Boden et al.(2009)]{boden09} Boden, A.~F., Akeson, 
R.~L., Sargent, A.~I., et al.\ 2009, \apjl, 696, L111 
\bibitem[Boden et al.(2012)]{boden12} Boden, A.~F., Torres, G., 
Duch{\^e}ne, G., et al.\ 2012, \apj, 747, 17 
\bibitem[Byrne(1986)]{byrne86} Byrne, P. B. 1986, Ir. Astron. J., 17, 294
\bibitem[D'Alessio et al.(1998)]{dalessio98} D'Alessio, P., Canto, J., Calvet, N., \& Lizano, S. 1998, \apj, 500, 411
\bibitem[D'Antona \& Mazzitelli(1998)]{dantona98} D'Antona, F., \& Mazzitelli, I. 1998, in ASP Conf.~Ser.~134, Brown Dwarfs and Extrasolar Planets, ed.~R.~Rebolo, E.~L.~Martin, \& M.~R.~Z. Osorio (San Francisco, CA: ASP), 442
\bibitem[D'Antona et al.(2000)]{dantona00} D'Antona, F., Ventura, P., \& Mazzitelli, I. 2000, \apj, 543, L77
\bibitem[Dartois et al.(2003)]{dartois03} Dartois, E., Dutrey, A., \& Guilloteau, S. 2003, \aap, 399, 773
\bibitem[Donati et al.(2011)]{donati11} Donati, J.-F., et al. 2011, \mnras, 417, 1747 
\bibitem[Duch{\^{e}}ne et al.(2006)]{duchene06} Duch{\^{e}}ne, G., Beust, H., Adjali, F., Konopacky, Q. M., \& Ghez, A. M. 2006, \aap, 457, L9
\bibitem[Dutrey et al.(1994)]{dutrey94} Dutrey, A., Guilloteau, S., \& Simon, M.\ 1994, \aap, 286, 149 
\bibitem[Dutrey et al.(1998)]{dutrey98} Dutrey, A., Guilloteau, S., Prato, L., Simon, M., Duvert, G., Schuster, K., \& M{\'{e}}nard, F. 1998, \aap, 338, L63 
\bibitem[Epchtein et al.(1997)]{epchtein97} Epchtein, N., et al. 1997, Messenger, 87, 27
\bibitem[Foreman-Mackey et al.(2012)]{foreman-mackay12} Foreman-Mackey, D., Hogg, D. W., Lang, D., \& Goodman, J. 2012, arXiv:1202.3665 
\bibitem[Gennaro et al.(2012)]{gennaro12} Gennaro, M., Prada Moroni, P. G., \& Tognelli, E. 2012, \mnras, 420, 986
\bibitem[Goodman \& Weare(2010)]{goodman10} Goodman, J., \& Weare, J. 2010, Comm. App. Math. Comp. Sci., 5, 65
\bibitem[Gom{\'e}z Maqueo Chew et al.(2012)]{gomezmaqueochew12} Gom{\'e}z Maqueo Chew, Y., et al. 2012, \apj, 745, 58
\bibitem[Guilloteau \& Dutrey(1998)]{guilloteau98} Guilloteau, S., \& Dutrey, A.\ 1998, \aap, 339, 467
\bibitem[Hartmann et al.(1998)]{hartmann98} Hartmann, L., Calvet, N., Gullbring, E., \& D'Alessio, P.\ 1998, \apj, 495, 385 
\bibitem[Hillenbrand \& White(2004)]{hillenbrand04} Hillenbrand, L. A., \& White, R. J. 2004, \apj, 604, 741
\bibitem[Ho et al.(2004)]{ho04} Ho, P. T. P., Moran, J. M., \& Lo, K. Y. 2004, \apj, 616, L1
\bibitem[Hoeting et al.(1999)]{hoeting99} Hoeting, J. A., Madigan, D., Raftery, A. E., \& Volinsky, C. T. 1999, Statist. Sci., 14, 4
\bibitem[Hughes et al.(2008)]{hughes08} Hughes, A. M., Wilner, D. J., Qi, C., \& Hogerheijde, M. R. 2008, \apj, 678, 1119
\bibitem[Hughes et al.(2011)]{hughes11} Hughes, A. M., Wilner, D. J., Andrews, S. M., Qi, C., \& Hogerheijde, M. R. 2011, \apj, 727, 85
\bibitem[Hut(1981)]{hut81} Hut, P. 1981, \aap, 99, 126
\bibitem[Hutchinson et al.(1990)]{hutchinson90} Hutchinson, M. G., Evans, A., Winkler, H., \& Spencer Jones, J. 1990, \aap, 234, 230 
\bibitem[Isella et al.(2007)]{isella07} Isella, A., Testi, L., Natta, A., Neri, R., Wilner, D., \& Qi, C. 2007, \aap, 469, 213
\bibitem[Jensen \& Mathieu(1997)]{jensen97} Jensen, E. L. N., \& Mathieu, R. D. 1997, \aj, 114, 301 
\bibitem[J{\o}rgensen \& Lindegren(2005)]{jorgensen05} J{\o}rgensen, B. R., \& Lindegren, L. 2005, \aap, 436, 127 
\bibitem[Kastner et al.(2008)]{kastner08} Kastner, J. H., Zuckerman, B., Hily-Blant, P., \& Forveille, T. 2008, \aap, 492, 469 
\bibitem[Kastner et al.(2011)]{kastner11} Kastner, J.~H., Sacco, 
G.~G., Montez, R., et al.\ 2011, \apjl, 740, L17 
\bibitem[Koerner et al.(1993)]{koerner93} Koerner, D. W., Sargent, A. I., \& Beckwith, S. V. W. 1993, Icarus, 106, 2
\bibitem[Lejeune et al.(1997)]{lejeune97} Lejeune, T., Cuisinier, F., \& Buser, R.\ 1997, \aaps, 125, 229 
\bibitem[Lynden-Bell \& Pringle(1974)]{lynden-bell74} Lynden-Bell, D., \& Pringle, J. E. 1974, \mnras, 168, 603
\bibitem[Mathieu et al.(1989)]{mathieu89} Mathieu, R. D., Walter, F. A., \& Myers, P. C. 1989, \aj, 98, 987
\bibitem[Mathieu et al.(1991)]{mathieu91} Mathieu, R. D., Adams, F. C., \& Latham, D. W. 1991, \aj, 101, 2184
\bibitem[Mathieu et al.(1997)]{mathieu97} Mathieu, R. D., et al. 1997, \aj, 113, 1841
\bibitem[Melo et al.(2001)]{melo01} Melo, C. H. F., Covino, E., Alcal{\'{a}}, J. M., \& Torres, G. 2001, \aap, 378, 898
\bibitem[Mendes et al.(1999)]{mendes99} Mendes, L. T. S., D'Antona, F., \& Mazzitelli, I. 1999, \aap, 341, 174
\bibitem[Morales-Calder{\'o}n et al.(2012)]{morales12} 
Morales-Calder{\'o}n, M., Stauffer, J.~R., Stassun, K.~G., et al.\ 2012, 
\apj, 753, 149 
\bibitem[Mundy et al.(1996)]{mundy96} Mundy, L. G., et al. 1996, \apj, 464, L169
\bibitem[{\"{O}}berg et al.(2011)]{oberg11} {\"{O}}berg, K. I., et al. 2011, \apj, 734, 98
\bibitem[Pavlyuchenkov et al.(2007)]{pavlyuchenkov07} Pavlyuchenkov, Y., Semenov, D., Henning, T., Guilloteau, S., Pi{\'{e}}tu, V., Launhardt, R., \& Dutrey, A. 2007, \apj, 669, 1262 
\bibitem[Pi{\'e}tu et al.(2007)]{pietu07} Pi{\'e}tu, V., Dutrey, A., \& Guilloteau, S.\ 2007, \aap, 467, 163 
\bibitem[Prato et al.(2002)]{prato02} Prato, L., Simon, M., Mazeh, T., Zucker, S., \& McLean, I. S. 2002, \apj, 579, L99 
\bibitem[Quast et al.(2000)]{quast00} Quast, G. R., Torres, C. A. O., de La Reza, R., da Silva, L., \& Mayor, M. 2000, IAU Symposium, 200, 28P 
\bibitem[Rodriguez et al.(2010)]{rodriguez10} Rodriguez, D. R., Kastner, J. H., Wilner, D., \& Qi, C. 2010, \apj, 720, 1684 
\bibitem[Schaefer et al.(2003)]{schaefer03} Schaefer, G. H., Simon, M., Nelan, E., \& Holfeltz, S. T. 2003, \aj, 126, 1971
\bibitem[Schaefer et al.(2006)]{schaefer06} Schaefer, G. H., Simon, M., Beck, T. L., Nelan, E., \& Prato, L. 2006, \aj, 132, 2618
\bibitem[Schaefer et al.(2008)]{schaefer08} Schaefer, G.~H., 
Simon, M., Prato, L., \& Barman, T.\ 2008, \aj, 135, 1659 
\bibitem[Schmidt-Kaler(1982)]{schmidt-kaler82} Schmidt-Kaler, T. 1982, in Land{\"o}lt Bornstein, Group VI, Vol. 2, ed. K.-H. Hellwege (Berlin:  Springer), 454
\bibitem[Sch{\"o}ier et al.(2005)]{scholier05} Sch{\"o}ier, F. L., van der Tak, F. F. S., van Dishoeck, E. F., \& Black, J. H. 2005, \aap, 432, 369 
\bibitem[Simon et al.(2000)]{simon00} Simon, M., Dutrey, A., \& Guilloteau, S. 2000, \apj, 545, 1034 
\bibitem[Siess et al.(1997)]{siess97b} Siess, L., Forestini, M., \& Dougados, C. 1997, \aap, 326, 1001
\bibitem[Siess \& Livio(1997)]{siess97a} Siess, L., \& Livio, M. 1997, \apj, 490, 785
\bibitem[Siess et al.(2000)]{siess00} Siess, L., Dufour, E., \& Forestini, M. 2000, \aap, 358, 593 
\bibitem[Skrutskie et al.(2006)]{skrutskie06} Skrutskie, M. F., et al. 2006, \aj, 131, 1163
\bibitem[Stassun et al.(2004)]{stassun04} Stassun, K. G., Mathieu, R. D., Vaz, L. P. R., Stroud, N., \& Vrba, F. J. 2004, \apjs, 151, 357
\bibitem[Stassun et al.(2012)]{stassun12} Stassun, K. G., Kratter, K. M., Scholz, A., \& Dupuy, T. J. 2012, \apj, in press (arXiv:1206.4930)
\bibitem[Steffen et al.(2001)]{steffen01} Steffen, A. T., et al. 2001, \aj, 122, 997
\bibitem[Stempels \& Gahm(2004)]{stempels04} Stempels, H. C., \& Gahm, G. F. 2004, \aap, 421, 1159 
\bibitem[Stempels(2012)]{stempels12} Stempels, H. C., 2012, in preparation
\bibitem[Strassmeier et al.(1993)]{strassmeier93} Strassmeier, K. G., Hall, D. S., Fekel, F. C., \& Scheck, M. 1993, \aaps, 100, 173
\bibitem[Tamazian et al.(2002)]{tamazian02} Tamazian, V. S., Docobo, J. A., White, R. J., \& Woitas, J. 2002, \apj, 578, 925
\bibitem[Tognelli et al.(2011)]{tognelli11} Tognelli, E., Prada Moroni, P. G., \& Degl'Innocenti, S. 2011, \aap, 533, A109 
\bibitem[Torres et al.(2006)]{torres06} Torres, C. A. O., Quast, G. R., da Silva, L., de la Reza, R., Melo, C. H. F., \& Sterzik, M. 2006, \aap, 460, 695
\bibitem[Torres et al.(2008)]{torres08} Torres, C.~A.~O., Quast, 
G.~R., Melo, C.~H.~F., 
\& Sterzik, M.~F.\ 2008, Handbook of Star Forming Regions, Volume II, 757 
\bibitem[Zacharias et al.(2010)]{zacharias10} Zacharias, N., et al. 2010, \aj, 139, 2184
\bibitem[Zahn(1977)]{zahn77} Zahn, J.-P. 1977, \aap, 57, 383
\bibitem[Zuckerman et al.(2001)]{zuckerman01} Zuckerman, B., Song, I., Bessell, M. S., \& Webb, R. A. 2001, \apj, 562, L87
\bibitem[Zuckerman \& Song(2004)]{zuckerman04} Zuckerman, B., \& Song, I.\ 2004, \araa, 42, 685 
\end{thebibliography}
\end{document}